\documentclass[10pt,aps,prd,twocolumn,showpacs,superscriptaddress,eqsecnum,nofootinbib]{revtex4}
\usepackage{amssymb}
\usepackage{xcolor}
\definecolor{navy}{RGB}{0,0,150}
\usepackage[colorlinks, linkcolor=navy, citecolor=navy, urlcolor=navy, plainpages=false, pdfstartview=FitH]{hyperref}
\usepackage{amsmath}
\usepackage{amsfonts}
\usepackage{mathrsfs}

\usepackage{amsmath,amssymb,amsfonts}
\usepackage{graphicx}

\def\be{\begin{equation}}
\def\ee{\end{equation}}
\def\ba{\begin{eqnarray}}
\def\ea{\end{eqnarray}}
\def\nn{\nonumber}

\newcommand{\ket}[1]{\vert{#1}\rangle} 

\newcommand{\ints}{{\int_\Sigma}} 

\newcommand{\Tr}{\mathrm{Tr}} 
\newcommand{\grav}{\mathrm{gr}} 
\newcommand{\kin}{\mathrm{kin}} 
\newcommand{\hil}{\mathcal{H}} 
\newcommand{\Euc}{H^{E}} 

\begin{document}


\title{ Canonical loop quantization of the lowest-order projectable Horava gravity}

\author{Xiangdong Zhang}
\affiliation{Department of Physics, South China University of
Technology, Guangzhou 510641, China}

\author{Jinsong Yang}
\affiliation{School of Physics, Guizhou University, Guiyang 550025, China}

\author{Yongge Ma}
\thanks{Corresponding author; mayg@bnu.edu.cn}
\affiliation{Department of Physics, Beijing Normal University, Beijing 100875, China}

\begin{abstract}
The Hamiltonian formulation of the lowest-order projectable Horava gravity, namely the so-called $\lambda$-$R$ gravity, is studied. Since a preferred
foliation has been chosen in projectable  Horava gravity, there is  no local Hamiltonian constraint in the theory. In contrast to general relativity, the constraint algebra of $\lambda$-$R$ gravity forms a Lie algebra. By canonical transformations, we further obtain the connection-dynamical formalism of the $\lambda$-R gravity theories with real $su(2)$-connections as configuration variables. This formalism enables us to extend the scheme of non-perturbative loop quantum gravity to the $\lambda$-$R$ gravity. While the quantum kinematical
framework is the same as that for general relativity, the Hamiltonian constraint operator of loop quantum $\lambda$-$R$ gravity can be well defined in the diffeomorphism-invariant Hilbert space. Moreover, by introducing a global dust degree of freedom to represent a dynamical time, a physical Hamiltonian operator with respect to the dust can be defined and the physical states satisfying all the constraints are obtained.

\end{abstract}

\pacs{04.60.Pp, 04.50.Kd}

\maketitle

\section{Introduction}
It is well known that all the fundamental interactions of the nature,
except for gravity, can be described in the framework of quantum field theory (QFT). Since gravity is universally coupled
to all the matter fields, the quantum nature of matter field imply that
gravity should be also quantized. In addition, around the singularities of the big bang and
black holes interior, the space-time curvature becomes divergent. Hence
it is generally expected that general relativity (GR), as a
classical theory, is no longer valid there, and quantum physics should
be taken into account. If a quantum theory of gravity could be available, the singularities would
be smoothed out by certain physically meaningful quantum description. Motivated by the above considerations, to realize the quantization of
gravity serves as one of the main driving forces in theoretical physics
in the past decades \cite{QG}, and various approaches have been
pursued, including string/M-Theory \cite{string} and loop quantum
gravity (LQG) \cite{Ro04,Th07,As04,Ma07}.

As a background independent approach to quantize GR, LQG has been widely investigated in the past 30 years \cite{Ro04,Th07,As04,Ma07}.
It is remarkable that, as a non-renormalizable theory, GR can be non-perturbatively
quantized by the loop quantization procedure. This background-independent quantization method relies on the key
observation that classical GR can be cast into the connection-dynamical formalism with the structure group of $SU(2)$. The LQG quantization method has been successfully generalized to $f(R)$ gravity \cite{Zh11,Zh11b}, scalar-tensor gravity \cite{ZhangMa11}, and Weyl gravity \cite{Ma18}.

The notion of time plays an important role in any quantum gravity theories and on how to implement
particular proposals in technical terms \cite{Isham94}. In the Hamiltonian framework of GR, one assumes that a Lorentzian spacetime
\textit{M} is diffeomorphic to a product
$\textit{M}={\mathbb R}\otimes\Sigma$
with $\Sigma$ being a smooth spacelike hypersurface, and ${\mathbb R}$ being a preferred time direction
following from the usual requirement of global hyperbolicity,
which ensures that the causal structure of spacetime is sufficiently
well behaved. The spacetime diffeomorphism invariance of GR in restored by the diffeomorphism and Hamiltonian constraints in the Hamiltonian framework. Thus, different choices of foliation can be considered as a part of
the gauge freedom of GR.

As a different kind of gravity theories, the so-called Ho\v{r}ava-Lifshitz gravity was proposed \cite{horava}, associated with a preferred
foliation of spacetime. As a consequence, these theories are only
invariant under a subset of spacetime diffeomorphisms, namely those that do not change the preferred foliation. The
remaining invariant group consists of three-dimensional diffeomorphisms acting independently on each leaf $\Sigma_t$
(labeled by time $t$) and space-independent time reparametrizations. The most general local action of the metric fields
which is at most quadratic in derivatives and invariant under this reduced symmetry group is not the concise Einstein-Hilbert
action, but in a rather complicated form \cite{horava}.

By giving up the space-time covariance, Ho\v{r}ava-Lifshitz gravity becomes renormalizable in  QFT perturbative quantization  \cite{Carlip12,Wang14,Wang16}. However, from the non-perturbative viewpoint, the LQG quantization method has not been extend to these theories. It is well known that the loop quantization highly relies on the connection-dynamical formalism of the corresponding gravity theories, while the connection-dynamical formalism of the Ho\v{r}ava-Lifshitz gravity is still absent. Note that due to  the extremely complicated
form of Ho\v{r}ava-Lifshitz gravity theories, one usually performs the quantization procedures in some simpler case, for examples, in lower dimensions \cite{Wang14,Wang16} or in the symmetry-reduced case such as the cosmological situations \cite{Pitelli16}.

The low energy limit of Ho\v{r}ava-Lifshitz gravity, which is suitable
for most astrophysical objects as well as cosmological applications \cite{Barausse11,Barausse13}, can be described by the following action
\begin{align}
S&=\frac{1-\beta}{16\pi G}\int dt\int_\Sigma d^3x N\sqrt{q}\left(K_{ab}K^{ab}-\frac{1+\nu}{1-\beta} K^2\right.\notag\\
&\hspace{2cm}\left.+\frac{1}{1-\beta}R+\frac{\sigma}{1-\beta}a_ia^i\right),
\end{align}
where $G$ is gravitational constant, $K_{ab}$ is the
extrinsic curvature of a spatial hypersurface $\Sigma$, $K\equiv
K_{ab}q^{ab}$, $R$ denotes the scalar curvature of the 3-metric
$q_{ab}$ induced on $\Sigma$, $a_i=\partial_i(\ln N)$, $\beta, \sigma$ and $\nu$ are coupling constants. The coupling constants must satisfy a series
of theoretical requirements, such as the absence of gradient instabilities
and ghosts \cite{BS11,JM04,GJ11}, as well as experimental constraints, including the absence of vacuum Cherenkov radiation \cite{Moore05}, solar system experiments \cite{Will14,BB15}, gravitational wave propagation bounds from GW170817 \cite{Sotiriou17,RB19}, and cosmological
constraints \cite{Yagi14,Yagi14L,Carroll04}. Those constraints suggest that $\beta$ and $\sigma$ are vanishingly small as  $\beta\leq 10^{-15}$ and $\sigma\leq 10^{-7}$.
However the other coupling constant $\nu$ is relatively unconstrained aside from the stability requirements
and cosmological bounds \cite{RB19,Carroll04,Afshordi09} such that $0\leqq \nu \lesssim0.01-0.1$.
Therefore, in this paper, we are going to quantize the four-dimensional simpler
model of gravity by setting $\beta=\sigma=0$ \cite{RB19,Barausse19}, which is the lowest-order Hor¡Šava gravity. This theory is sometimes called as $\lambda$-$R$ gravity model \cite{Kiefer94,BR12,Loll14,Loll17}. Thus the action of $\lambda$-$R$ gravity reads \cite{Kiefer94,BR12,Loll14,Loll17}
\begin{align}
S&=\frac{1}{16\pi G}\int dt\ints d^3x\sqrt{q}N(K_{ab}K^{ab}-\lambda K^2+R)\notag\\
&\equiv\int d^4x \mathcal{L}\label{lambdaR}
\end{align}
with the coupling parameter $\lambda\equiv1+\nu$. This theory serves as the minimal
generalization of GR, since action \eqref{lambdaR} reduces to Einstein-Hilbert action by setting $\lambda=1$. It was first proposed and investigated in a purely classical context in Ref. \cite{Kiefer94}. Though it is simpler, the $\lambda$-$R$ gravity theory shares the same kinetic term and the symmetry of the Ho\v{r}ava-Lifshitz gravity. It has been shown in Refs. \cite{BR12,Loll14,Loll17},  that the nonprojectable $\lambda$-$R$ gravity models are equivalent
to GR in the asymptotically flat case, while the projectable sector of $\lambda$-$R$ gravity is inequivalent to GR. More precisely, by choosing a preferred foliation the usual local Hamiltonian constraint of GR was removed. As shown in Refs. \cite{Kobakhidze10,Blas09}, the absence of the local constraint leads to an additional strongly coupled scalar
degree of freedom, which becomes dynamical
here. Then the coupling of $\lambda$-$R$
gravity to matter would suggest a universal scalar (fifth) force in
nature, which has not been seen. Nevertheless, the projectable theory provides a practicable model to test the scheme of LQG. Thus, we will focus on
the projectable model of $\lambda$-$R$ gravity, where the lapse function $N$ is only a function of time $t$ \cite{Loll14,Loll17}.

This paper is organized as follows: We will present a detailed Hamiltonian analysis of $\lambda$-$R$ gravity to obtain its connection-dynamical formalism in section \ref{Section2}.
Then in section \ref{Section3}, the $\lambda$-$R$ gravity will be non-perturbatively quantized by the LQG method based on the connection dynamics, and the quantum Hamiltonian constraint operator for $\lambda$-$R$ gravity will be constructed. In  section \ref{Section4}, the non-rotational dust field will be introduced to represent a dynamical time and the physical Hamiltonian operator will be defined so that  the physical states can be obtained.
Our result will be summarized in the last section. Throughout the paper, we use Latin alphabet $a,b,c,\cdots$ for spatial indices, and
$i,j,k,\cdots$ for internal indices, and set $8\pi G=1$ for simplicity.

\section{Hamiltonian analysis}\label{Section2}

Starting from action \eqref{lambdaR}, by Legendre transformation, the
momentum conjugate to the dynamical variable $q_{ab}$ reads
\begin{align}
p^{ab}&=\frac{\partial\mathcal
{L}}{\partial\dot{q}_{ab}}=\frac{N\sqrt{q}}{2}(K^{ab}-\lambda
Kq^{ab}). \label{04}
\end{align}
The Hamiltonian of $\lambda$-$R$ gravity can be derived as a liner combination of constraints \cite{Kobakhidze10,BR12,Loll14},
\begin{align}
H_{total}=\int_\Sigma d^3x(N^aC_a+NC),\label{htotal}
\end{align}
where the shift vector $N^a$ is a vector-valued function on $\Sigma$, $N$ is a constant in every spatial slice. The smeared diffeomorphism and Hamiltonian constraints read
respectively
\begin{align}
C(\overrightarrow{N})&=\int_\Sigma d^3xN^aC_a \equiv\int_\Sigma
d^3xN^a\left(-2D^b(p_{ab})\right),\label{dc}\\
\tilde{C}_0&=\int_\Sigma d^3xC \nn\\
&\equiv\int_\Sigma
d^3x\left(\frac2{\sqrt{q}}\left(p_{ab}p^{ab}-\frac{\lambda}{3\lambda-1}p^2\right)-\frac12\sqrt{q}R\right)\notag\\
&\hspace{4.8cm}\left.\right.\label{hc}
\end{align}
where we fix $N=1$ from now on. Note that the Hamiltonian constraint $\tilde{C}_0$ is a global constraint rather than a local one, which does not generate local gauge transformations. The symplectic structure is given by the following non-trivial Poisson bracket between the canonical variables,
\begin{align}
\{q_{ab}(x),p^{cd}(y)\}&=\delta^{(c}_a\delta^{d)}_b\delta^3(x,y).\label{poission}
\end{align}
Straightforward calculations show that the constraints (\ref{dc}) and (\ref{hc}) comprise a first-class system as\cite{Kobakhidze10}:
\begin{align}
\{C(\overrightarrow{N}),C(\overrightarrow{N}^\prime)\}&=C([\overrightarrow{N},\overrightarrow{N}^\prime]), \\
\{\tilde{C}_0,C(\overrightarrow{N})\}&=0, \\
\{\tilde{C}_0,\tilde{C}_0\}&=0.
\end{align}
These constraint algebra has the nice property of a Lie algebra, and the diffeomorphism
constraints also nicely form an ideal. This implies that in the canonical quantization it is possible to define the Hamiltonian constraint operator
directly on the diffeomorphism invariant Hilbert space.

To set up the classical foundation of loop quantization, we can employ the canonical transformation technique for metric theories of gravity to obtain the connection dynamical formalism of $\lambda$-$R$ gravity. Let
\begin{align}
\tilde{K}^{ab}=K^{ab}-\frac{1-\lambda}{2}Kq^{ab}.
\end{align} Then the conjugate momentum $p^{ab}$ of $q_{ab}$ could be rewritten as
\begin{align}
p^{ab}=\frac{\sqrt{q}}{2}(\tilde{K}^{ab}-
\tilde{K}q^{ab}).
\end{align}
We define the new geometric variables
through
\begin{align}
E^a_i=\sqrt{q}e^a_i,\quad
\tilde{K}^a_i\equiv\tilde{K}^{ab}e_b^j\delta_{ij},
\end{align}
where $e^a_i$ is the triad on $\Sigma$ such that $q_{ab}e^a_ie^b_j=\delta_{ij}$.
Now we extend the phase space of the theory to the space consisting of pairs $(E^a_i, \tilde{K}_a^i)$. It is then easy to see that the
symplectic  structure (\ref{poission}) can be derived from the following
Poisson brackets:
\begin{align}
\{\tilde{K}^j_a(x),E_k^b(y)\}&=-\delta^b_a\delta^j_k\delta^3(x,y),\\
\{E^a_j(x),E^b_k(y)\}&=0,\\
\{\tilde{K}_a^j(x),\tilde{K}_b^k(y)\}&=0.
\end{align}
Thus there is a direct symplectic reduction from the extended phase space to the original one. In this sense the transformation from conjugate pairs $(q_{ab},p^{cd})$ to
$(E^a_i,\tilde{K}^j_b)$ is canonical. Note that the symmetry of $\tilde{K}^{ab}$, i.e.
$\tilde{K}^{ab}=\tilde{K}^{ba}$, gives rise to an additional constraint in the extend phase space as:
\begin{align}
G_{jk}\equiv\tilde{K}_{a[j}E^a_{k]}=0. \label{gaussian}
\end{align}
So we can make a second canonical transformation by defining \cite{Th07,Ma07}:
\begin{align}
A^i_a=\Gamma^i_a+\gamma\tilde{K}^i_a,
\end{align}
where $\Gamma^i_a$ is the spin connection determined by the densitized triad $E^a_i$, and $\gamma$ is a nonzero
real number which is usually called as Barbero-Immirzi parameter in the community of LQG \cite{Ba}. It is clear that our new variable $A^i_a$ coincides with the Ashtekar-Barbero connection of GR \cite{As86,Ba} when $\lambda=1$. Therefore our new variable $A^i_a$ serves as an extension of the Ashtekar-Barbero connection for $\lambda$-$R$ gravity.
The Poisson brackets among the new variables read:
\begin{align}
\{A^j_a(x),E_k^b(y)\}&=\gamma\delta^b_a\delta^j_k\delta(x,y),\\
\{A_a^i(x),A_b^j(y)\}&=0, \\
\{E_j^a(x),E_k^b(y)\}&=0.
\end{align}
Now, the phase space of $\lambda$-$R$ gravity consists of conjugate pairs $(A_a^i,E^b_j)$. Combining
Eq. (\ref{gaussian}) with the compatibility condition:
\begin{align}
\partial_aE^a_i+\epsilon_{ijk}\Gamma^j_aE^{ak}=0,
\end{align}
we obtain the standard Gaussian constraint
\begin{align}
\mathcal
{G}_i=\mathscr{D}_aE^a_i\equiv\partial_aE^a_i+\epsilon_{ijk}A^j_aE^{ak},
\label{GC}
\end{align}
which justifies $A^i_a$ as an $su(2)$-connection. Note
that, had we let $\gamma=\pm i$, the (anti-)self-dual complex
connection formalism would be obtained. The original diffeomorphism
constraint as well as the Hamiltonian constraint can be expressed in terms of new variables up to Gaussian
constraint as
\begin{align}
C^{\lambda R}_a &=\frac1\gamma F^i_{ab}E^b_i=0,  \\
C_0&=\ints d^3x C^{\lambda R}\nn\\
&=\frac{1}{2}\int_\Sigma d^3x\left(\left(F^j_{ab}-(1+\gamma^2)\epsilon_{jmn}\tilde{K}^m_a\tilde{K}^n_b\right)
\frac{\epsilon_{jkl} E^a_kE^b_l}{\sqrt{q}}\right.\notag\\
&\hspace{2cm}\left.+\frac{2-2\lambda}{1-3\lambda}\frac{(\tilde{K}^i_aE^a_i)^2}{\sqrt{q}}\right)=0,
\label{hamilton}
\end{align}
where $F^i_{ab}\equiv2\partial_{[a}A^i_{b]}+\epsilon^i_{kl}A_a^kA_b^l$ is
the curvature of the $su(2)$-connection $A_a^i$. The total Hamiltonian can be expressed as a linear combination
\begin{align}
H_{total}=\ints d^3x\left(\Lambda^i\mathcal {G}_i+N^aC^{\lambda R}_a+C^{\lambda R}\right).
\end{align}
It is easy to check that the  smeared Gaussian constraint, $\mathcal
{G}(\Lambda):=\int_\Sigma d^3x\Lambda^i(x)\mathcal {G}_i(x)$,
generates $SU(2)$ gauge transformations on the phase space, while
the smeared constraint $\mathcal
{V}(\overrightarrow{N}):=\int_\Sigma d^3xN^a(C^{\lambda R}_a-A_a^i\mathcal
{G}_i)$ generates spatial diffeomorphism transformations on the
phase space. Together with the Hamiltonian constraint, it is straightforward to show that the constraints algebra
has the following form:
\begin{align}
\{\mathcal {G}(\Lambda),\mathcal
{G}(\Lambda^\prime)\}&=\mathcal
{G}([\Lambda,\Lambda^\prime]),\label{eqsA} \\
\{\mathcal
{G}(\Lambda),\mathcal{V}(\overrightarrow{N})\}&=-\mathcal{G}(\mathcal
{L}_{\overrightarrow{N}}\Lambda),\\
\{\mathcal {G}(\Lambda),C_0\}&=0,\\
\{\mathcal {V}(\overrightarrow{N}),\mathcal
{V}(\overrightarrow{N}^\prime)\}&=\mathcal
{V}([\overrightarrow{N},\overrightarrow{N}^\prime]), \\
\{\mathcal {V}(\overrightarrow{N}),C_0\}&=0,\label{eqsE}\\
\{C_0,C_0\}&=0.\label{eqsb}
\end{align}
Hence the constraints are all of first class. To summarize, the $\lambda$-$R$ gravity have
been cast into the $su(2)$-connection dynamical formalism. It is worth noting that in the LQG of GR, although the Hamiltonian constraint is well defined in gauge invariant Hilbert space $\hil_{G}$, it is difficult to define it
directly in the diffeomorphism invariant Hilbert space $\hil_{Diff}$. Moreover, since the constraint algebra
of GR does not form a Lie algebra, the quantum anomaly might appear after quantization. In contrast, the diffeomorphism
constraints nicely form an ideal in $\lambda$-$R$ gravity. Therefore the Hamiltonian constraint operator could be defined
directly in $\hil_{Diff}$.

\section{Quantization of $\lambda$-$R$ theory} \label{Section3}

Based on the connection dynamical formalism, the
nonperturbative loop quantization procedure can be straightforwardly
extended to the $\lambda$-$R$ gravity. The kinematical structure of $\lambda$-$R$ gravity is just the same as
that of LQG for GR \cite{As04,Ma07}. The kinematical Hilbert space, $\hil_\kin:=\hil^\grav_\kin$, of the $\lambda$-$R$ gravity is spanned by the spin-network basis ${\cal \psi}_{\alpha}(A)=\ket{\alpha,j,i}$ over graphs $\alpha\subset\Sigma$, where $j$ labels the irreducible representations of $SU(2)$ associated to the edges of $\alpha$ and $i$ denotes the intertwiners assigned to the vertices linking the edges. The basic operators are
the quantum analogue of holonomies, $h_e(A)=\mathcal {P}\exp\int_eA_a$, of connections and densitized triads smeared over
2-surfaces, $E(S,f):=\int_S\epsilon_{abc}E^a_if^i$. Note that the whole construction is background independent, and the
spatial geometric operators of LQG, such as the area \cite{Ro95}, the volume \cite{As97,Ma16} and the length operators \cite{Th98,Ma10}, are
still valid here. As in LQG, it is straightforward to promote the Gaussian constraint $\mathcal {G}(\Lambda)$ to a well-defined
operator \cite{Th07,Ma07}. It's kernel is the internal gauge invariant Hilbert space $\mathcal {H}_G$ with gauge invariant
spin-network basis. Moreover the diffeomorphisms of $\Sigma$ act covariantly on the cylindrical functions in $\mathcal {H}_G$, and hence the
so-called group averaging technique can be employed to solve the
diffeomorphism constraint \cite{As04,Ma07}, which gives rise to the desired gauge and diffeomorphism invariant Hilbert space $\mathcal {H}_{Diff}$ for the $\lambda$-$R$ gravity.

The remaining nontrivial task for $\lambda$-$R$ gravity is to implement the Hamiltonian constraint (\ref{hamilton}) at
quantum level. In order to compare the Hamiltonian constraint of $\lambda$-$R$ gravity
with that of GR in connection formalism, we write Eq. (\ref{hamilton}) as $C_0=\sum^3_{i=1}C_i$, where
the terms $C_1,C_2$ take the same form as the Euclidean and Lorentzian terms in GR \cite{As04,Ma07}, i.e., \ba
C_1&=&H^E(1)=\frac{1}{2}\int_\Sigma d^3xF^j_{ab}\frac{\epsilon_{jkl} E^a_kE^b_l}{\sqrt{q}},\\
C_2&=&-\frac{(1+\gamma^2)}{2}\int_\Sigma d^3x\epsilon_{jmn}\tilde{K}^m_a\tilde{K}^n_b
\frac{\epsilon_{jkl} E^a_kE^b_l}{\sqrt{q}}.
\ea  Hence the difference comes from the completely new term,
\begin{align}
C_3=\int_\Sigma d^3x\frac{(2-2\lambda)}{1-3\lambda}\frac{(\tilde{K}^i_aE^a_i)^2}{\sqrt{q}}.\label{c3}
\end{align}
This term can be treated by the similar regularization techniques developed for the Hamiltonian in the LQG \cite{Th07}. We may triangulate $\Sigma$ in adaptation to some graph $\alpha$ underling a cylindrical function in $\hil_\kin$ and reexpress connections by holonomies. To this aim, we first note the following classical identity
\begin{align}
\tilde{K}=\ints d^3x\tilde{K}^i_aE^a_i=\frac{1}{\gamma^2}\{H^E(1),V\},
\end{align}
where $H^E(1)$ is the Euclidean term and $V$ is the volume \cite{Th07}. Therefore, one can further regularize Eq. \eqref{c3}
by the point-splitting method and obtain
\begin{align}
C_3&=\lim_{\epsilon\rightarrow 0}C^\epsilon_3=\lim_{\epsilon\rightarrow 0}\int_\Sigma d^3y\int_\Sigma
d^3x\frac{(2-2\lambda)}{1-3\lambda}\chi_\epsilon(x-y)\notag\\
&\hspace{3cm}\times\frac{\tilde{K}^i_a(x)E^a_i(x)}{\sqrt{V_{U^\epsilon_x}}}\frac{\tilde{K}^j_b(y)E^b_j(y)}{\sqrt{V_{U^\epsilon_y}}},\label{c3regularize}
\end{align}
where $\chi_\epsilon(x-y)$ is the characteristic function of a box $U^\epsilon_x$ containing $x$ with scale $\epsilon$ and
satisfies the relation
\begin{align}
\lim_{\epsilon\rightarrow
0}\frac{\chi_\epsilon(x-y)}{\epsilon^3}=\delta^3(x-y),
\end{align}
and $V_{U^\epsilon_x}$ denotes the volume of $U^\epsilon_x$. Now, we triangulate $\Sigma$ into elementary
tetrahedra $\Delta$ with scale $\epsilon$, and denote the triangulation by ${\cal T}$. For each $\Delta$, we single out one of its vertices, and call it $v(\Delta)$. Then, as $\Delta\rightarrow v(\Delta)$, we have
\begin{align}
\int_\Delta d^3x\frac{\tilde{K}^i_a(x)E^a_i(x)}{\sqrt{V_{U^\epsilon_x}}}\approx\frac{2}{\gamma^2}\left\{\Euc_\Delta,\sqrt{V_{U_{v(\Delta)}^\epsilon}}\right\},
\end{align}
where
\begin{align}
H^E_\Delta=\frac{2}{3\gamma}\,\epsilon^{IJK}{\rm Tr}\left(h_{\alpha_{IJ}(\Delta)}h_{s_K(\Delta)}\left\{h^{-1}_{s_K(\Delta)},V_{U_v^\epsilon}\right\}\right).
\end{align}
Here $s_I(\Delta), I=1,2,3,$ denote the three edges of $\Delta$ incident at $v(\Delta)$, $(I,J,K)\in\{(1,2,3),(2,3,1),(3,1,2)\}$ such that the triple $(s_I(\Delta),s_J(\Delta),s_K(\Delta))$ has positive orientation induced by $\Sigma$, and $\alpha_{IJ}(\Delta):=s_I(\Delta)\circ a_{IJ}(\Delta)\circ s_J(\Delta)$ is the loop based at $v(\Delta)$ with $a_{IJ}(\Delta)$ being the edge of $\Delta$ connecting those endpoints of $s_I(\Delta)$ and $s_J(\Delta)$ which are distinct from $v(\Delta)$. Thus $C^\epsilon_3$ in Eq. \eqref{c3regularize} can be expressed as
\begin{align}\label{c3regularize-ex}
C^\epsilon_3&=\frac{4}{\gamma^4}\frac{(2-2\lambda)}{1-3\lambda}\sum_{\Delta,\Delta'\in {\cal T}}\chi_\epsilon(v(\Delta)-v(\Delta'))\notag\\
&\hspace{0.3cm}\times\left\{H^E_\Delta,\sqrt{V_{U_{v(\Delta)}^\epsilon}}\right\}\left\{H^E_{\Delta'},\sqrt{V_{U_{v(\Delta')}^\epsilon}}\right\}.
\end{align}
Note that all the terms in \eqref{c3regularize-ex} including the Euclidean term  $H^E_\Delta$ and volume $V_{U_{v(\Delta)}^\epsilon}$ could be promoted as well-defined operators in the gauge-invariant Hilbert space $\mathcal {H}_G$. Furthermore, for a given graph $\alpha$, one constructs a triangulation ${\cal T}(\alpha)$ of $\Sigma$ adapted to $\alpha$ \cite{Th07}. Notice that the volume operator acts only at vertices of $\alpha$, and for sufficiently small $\epsilon$ the function $\chi_\epsilon(v(\Delta),v(\Delta'))=0$ unless $v(\Delta)=v(\Delta')$. Thus \eqref{c3regularize-ex} can also be promoted as a well-defined regularized operator acting on any $\psi_\alpha(A)\in\mathcal {H}_G$ as:
\begin{align}
\hat{C}^\epsilon_3\,\psi_\alpha(A)&=\frac{4}{\gamma^4(i\hbar)^2}\frac{(2-2\lambda)}{1-3\lambda}\sum_{v\in V(\alpha)}\frac{8^2}{E(v)^2}\notag\\
&\hspace{0.4cm}\times\sum_{v(\Delta)=v(\Delta')=v}\left[\hat{H}^E_\Delta,\sqrt{\hat{V}_v}\right]\notag\\
&\hspace{2.4cm}\times\left[\hat{H}^E_{\Delta'},\sqrt{\hat{V}_v}\right]\psi_\alpha(A),
\end{align}
where the first summation is over the vertices $v$ of $\alpha$, the second summation is over $\Delta$ with $v(\Delta)=v$, $E(v)=\dbinom{n(v)}{3}$ is the possible choices of triples for a vertex $v$ with $n(v)$ edges, and
\begin{align}
\hat{H}^E_\Delta:=\frac{2}{3i\hbar\gamma}\epsilon^{IJK}\Tr\left(\hat{h}_{\alpha_{IJ}(\Delta)}\hat{h}_{s_K(\Delta)}[\hat{h}^{-1}_{s_K(\Delta)},\hat{V}_v]\right).
\end{align}
In LQG of GR, because the diffeomorphism-invariant Hilbert space $\hil_{Diff}$ is not preserved by the Hamiltonian constraint operator, the Hamiltonian operator can only be well defined in $\mathcal {H}_G$ rather than $\hil_{Diff}$. However, in $\lambda$-$R$ gravity, since the lapse $N$ is a constant, $\hil_{Diff}$ would be preserved by the Hamiltonian constraint operator, and hence we can further define the Hamiltonian operator in $\hil_{Diff}$. Note that a diffeomorphism-invariant state can be produced from a state $\psi_\alpha(A)\in\mathcal {H}_G$ by the group averaging method as \cite{As04,Th07,Ma07}
\begin{align}
\hat{P}_{Diff_\alpha}\psi_\alpha(A):=\frac{1}{n_\alpha}\sum_{\varphi\in GS_\alpha}\hat{U}_\varphi \psi_\alpha(A)\label{diffinvariant} ,
\end{align}
where the operator $\hat{U}_\varphi$ denotes the finite diffeomorphism  $\varphi: \Sigma\rightarrow \Sigma$,
$GS_\alpha=Diff_\alpha/TDiff_\alpha$ is the group of graph symmetries with $Diff_\alpha$ being the group of all diffeomorphisms preserving the
graph $\alpha$, $TDiff_\alpha$ is its subgroup which has trivial action on $\alpha$, and $n_\alpha$ is the number of the elements in $GS_\alpha$.

Since the regularized operator $\hat{C}^\epsilon_3$  with different value of $\epsilon$ are diffeomorphic to each other, we can naturally define the action of the limit operator $\hat{C_3}=\lim_{\epsilon\rightarrow
0}\hat{C^\epsilon_3}$ on the diffeomorphism-invariant state as
\begin{align}
\hat{C_3}\hat{P}_{Diff_\alpha}\psi_\alpha(A):=\lim_{\epsilon\rightarrow
0}\frac{1}{n_{\alpha(\epsilon)}}\sum_{\varphi\in GS_{\alpha(\epsilon)}}\hat{U}_{\varphi } \hat{C^\epsilon_3}\psi_\alpha(A),
\label{C3action}
\end{align}
where $\alpha(\epsilon)$ represents the new graphs produced by the action of $\hat{C^\epsilon_3}$ on $\alpha$. Note that Eq. \eqref{C3action} does not depend on $\epsilon$, since all the graphs $\alpha(\epsilon)$ are diffeomorphism equivalent to each other. Similar to the definition of $\hat{C_3}$, it is straightforward to define the whole Hamiltonian constraint operator $\hat{C}_0$ in $\hil_{Diff}$ as
\begin{align}
\hat{C}_0\hat{P}_{Diff_\alpha}\psi_\alpha(A)&:=\lim_{\epsilon\rightarrow
0}\frac{1}{n_{\alpha(\epsilon)}}\sum_{\varphi\in GS_{\alpha(\epsilon)}}\sum_{i=1,2,3}\hat{U}_{\varphi }\notag\\
&\hspace{2.3cm}\times\hat{C^\epsilon_i}\,\psi_\alpha(A), \label{CNDiff}
\end{align}
with
\begin{align}
\hat{C}^\epsilon_1&=\sum_{v\in V(\alpha)}\frac{8}{E(v)}\sum_{v(\Delta)=v}\hat{H}^E_\Delta,\\
\hat{C}^\epsilon_2&=-\frac{4(1+\gamma^2)}{3(i\hbar\gamma)^3}\sum_{v\in V(\alpha)}\frac{8}{E(v)}\sum_{v(\Delta)=v}\epsilon^{IJK}\notag\\
&\hspace{0.8cm}\times\Tr\left(\hat{h}_{s_I(\Delta)}[\hat{h}^{-1}_{s_I(\Delta)},\hat{\tilde{K}}_v]\hat{h}_{s_J(\Delta)}[\hat{h}^{-1}_{s_J(\Delta)},\hat{\tilde{K}}_v]\right.\notag\\
&\hspace{1.9cm}\times\left.\hat{h}_{s_K(\Delta)}[\hat{h}^{-1}_{s_K(\Delta)},\hat{V}_v]\right),
\end{align}
where $\hat{\tilde{K}}_v:=\frac{1}{i\hbar\gamma^{2}}[\hat{H}^E_v,\hat{V}_v]$ with $\hat{H}^E_v:=\sum_{v(\Delta)=v}\hat{H}^E_\Delta$. Note that, to have a well-defined adjoint operator of $\hat{C}_0$ \cite{Ma15}, we used the freedom of choosing the spin representations attached to each new added loop in \eqref{CNDiff} to ensure that the valence of any vertex would not be changed by the action of $\hat{C}_0$.

\section{A Physical Hamitonian and physical states}\label{Section4}

It should be noted that even in projectable $\lambda$-$R$ gravity, due to the existence of the global Hamiltonian constraint, there still exists a global gauge freedom corresponding to the global time reparametrization. Thus, in the corresponding quantum theory, the Hamiltonian  constraint operator has to vanish on physical states. Therefore, the time problem of quantum gravity is still there. The purpose of this section  is to overcome this problem by introducing a single
global dust degree of freedom to represent a dynamical time. In a theory of gravity with time reparametrization invariance, in order to pick up a unique time to represent the evolution of physical states\cite{Isham94}, one naturally takes the viewpoint of relational evolution \cite{Lewandowski10,Lewandowski15,RS94,Kuchar91}. This
allows one to map the totally constrained theory into a theory
with a true nonvanishing Hamiltonian with respect to some chosen dynamical (emergent) time variable. The dynamical "time" can be achieved at the classical level as well as the quantum level. The combination of LQG with the relational evolution
framework makes it possible to solve the quantum Hamiltonian constraint.

The action of non-rotational dust model in a covariant spacetime reads
\ba
S=-\frac{1}{2}\int d^4x\sqrt{-g} M(g^{ab}\partial_aT\partial_bT+1),
\ea where $T$ is the configuration variable of the non-rotational dust, and $M$ is the rest mass density of the dust field. Its Hamiltonian can be written as \cite{Pawlowski12}
\ba
H_{D}=\int d^3x\left[N\sqrt{\pi^2+q^{ab}C^D_aC^D_b}+N^aC^D_a\right],\nn\\
\ea where $\pi$ is the conjugate momentum of $T$ and $C^D_a=-\pi\partial_aT$. In order to introduce the non-rotational dust model, which was widely used in LQG literatures \cite{Pawlowski12,Thiemann15,Lewandowski17,Husain15}, to represent a dynamical time for the $\lambda$-$R$ gravity, we consider the case that the dust is adapted to the spacetimes of Horava gravity so that the time foliation of the spacetimes coincides with the hypersurfaces of constant $T$. In the other words, certain function $t(T)$ of the dust configuration variable $T$ is employed to define the given time foliation of Horava spacetimes. Note that $t(T)$ needs not to be a fixed function. Thus the global time
reparametrization freedom still exists.

As the gauge group of Horava theory consists of the foliation-preserving diffeomorphisms, the projectable version of the theory concerns only the case  that the lapse function depends only on time $t$ \cite{Loll14}.
Note that for the adapted non-rotational dust, we have $q^{ab}\partial_aT=0$, and hence
the dust has no local degrees of freedom.  Thus, in the case of projectable $\lambda$-$R$ gravity with the adapted non-rotational dust, the global Hamiltonian constraint reads
\begin{align}
C_{total}&=\int d^3x(\pi(x)+h(x))\notag\\
&:=\int d^3x\left(\pi(x)+C^{\lambda R}(x)\right)=0.\label{HamiltonGD1}
\end{align}
Hence one can define a physical Hamiltonian $h_{phy}=\int d^3x h(x)$ which generates the evolution of the system with respect to the dynamical "time" $T$.

In the quantum theory, one would expect to implement the constraint corresponding to \eqref{HamiltonGD1} through a Schrodinger-like equation
\begin{align}
i\hbar\frac{\partial}{\partial T}\Phi(A,T)=\hat{h}_{phy}\Phi(A,T)\label{Schrodingereqgr}
\end{align}
for certain quantum states $\Phi(A, T)$. Note that in certain simplified models of quantum gravity, there are different ideas  to treat the Hamiltonian constraint as a true Hamiltonian \cite{Gryb18,Carlini97}. Since loop quantum $\lambda$-$R$ gravity has been constructed in previous sections and the gravitational Hamiltonian constraint $\hat{C}_0$ is well defined by \eqref{CNDiff} on any diffeomorphism-invariant state $\Phi_{[\alpha]} (A)=\hat{P}_{Diff_\alpha}\psi_\alpha(A)\in\hil_{Diff}$, it is convenient to define the physical Hamiltonian operator $\hat{h}_{phy}$ as a self-adjoint extension of the symmetric operator $\frac{1}{2}(\hat{C}_0+\hat{C}_0{}^{\dag})$. Then the general solutions to Eq. \eqref{Schrodingereqgr} read
\begin{align}
\Phi_{[\alpha']} (A,T)=e^{-\frac{i}{\hbar}\hat{h}_{phy}T}\Phi_{[\alpha]} (A),
\end{align}
with an arbitrary given $\Phi_{[\alpha]} (A)\in \hil_{Diff}$. Thus, the physical Hilbert space of the coupled system is unitarily isomorphic to $\hil_{Diff}$.

\section{Conclusion}\label{Section6}

In the previous sections, a detailed construction of connection-dynamical formalism of the lowest-order projectable Horava gravity is given. This theory is the so-called $\lambda$-$R$ gravity. Since a preferred foliation has been chosen in projectable Horava gravity, there is no local Hamiltonian constraint. We obtain a connection-dynamics with real
$su(2)$-connections as configuration variables. In contrast to GR, the constraint algebra of $\lambda$-$R$ gravity forms a Lie algebra, and the Hamiltonian (\ref{hamilton}) possess an extra term which would vanish for $\lambda=1$. This classical connection-dynamical formalism enables us to extend the scheme of non-perturbative loop quantum gravity to the $\lambda$-$R$ theories of gravity. While the quantum kinematical framework is the same as that for GR, the global Hamiltonian constraint of loop quantum $\lambda$-$R$ gravity has been rigorously constructed as a well-defined operator in the diffeopmorphism-invariant Hilbert space.

To overcome the time problem related to the global time reparamatrization freedom of the projectable $\lambda$-$R$ gravity, the non-rotating dust adapted to the Horava spacetimes is introduced as a dynamical time. The physical time evolution with respect to the dust is then naturally defined. As a result, the quantum dynamics of the coupled system is dictated by a Schrodinger-like equation. For an arbitrarily given initial diffeomorphism-invariant state, the physical quantum Hamiltonian operator would generate and thus completely determine the forthcoming quantum state with respect to the dynamical time. Moreover, the physical states we obtained satisfy all the constraints, and the physical Hilbert space of the coupled system is unitarily isomorphic to the diffeomorphism-invariant Hilbert space of $\lambda$-$R$ gravity. Therefore, we obtained a quantum theory of gravity in which the Dirac algorithm of canonical quantization for a totally constrained system could be completely realized.

There are of course a few issues that deserves further investigating in our loop quantum $\lambda$-$R$ theory of gravity. First, it is interesting to study some symmetry-reduced models of our loop quantum $\lambda$-$R$ gravity, which might tell us more physical properties of the quantum $\lambda$-$R$ gravity. Second, how to extend LQG to the non-projectable version of $\lambda$-$R$ gravity is an interesting issue. Third, if our result could be generalized to the general Ho\v{r}ava-Lifshitz gravity, it would be helpful to get a better understanding on the quantum gravity without Lorentzian invariance.

\begin{acknowledgments}
We would like to thank the anonymous referee for the constructive comments and suggestions to the initial version of our manuscript. This work is supported by NSFC with Grants No. 11775082, No. 11875006, No. 11961131013, and No. 11765006.
\end{acknowledgments}


%

\end{document}